\documentstyle[prl,aps,draft,twocolumn]{revtex}
\begin{document}
\draft

\title{An experimental measurement of the staggered magnetization curve
for a Haldane spin chain.}

\author{A. Zheludev$^{(1)}$, E. Ressouche$^{(2)}$, S. Maslov$^{(1)}$, T. Yokoo$^{(1,3)}$,
 S. Raymond$^{(1,2)}$,  and J. Akimitsu$^{(3)}$}
\address{(1) Physics Department, Brookhaven National Laboratory, Upton,
New York 11973 USA.
 (2) DRFMC/SPSMS/MDN Centre d'Etudes Nucleaires, 17 rue des Martyres, 38054
 Grenoble Cedex, France.
 (3)Department of
Physics, Aoyama-Gakuin University, 6-16-1, Chitosedai, Setagaya-ku,
Tokyo 157 Japan. }
\date{\today}

\maketitle
\begin{abstract}
Long-range magnetic ordering in $R_{2}$BaNiO$_{5}$ ($R$=magnetic rare
earth) quasi-1-dimensional mixed-spin antiferromagnets is described by a
simple mean-field model that is based on the intrinsic staggered
magnetization function of isolated Haldane spin chains for the
Ni-subsystem, and single-ion magnetization functions for the rare earth
ions. The model is applied to new experimental results obtained in
powder diffraction experiments on Nd$_{2}$BaNiO$_{5}$ and
NdYBaNiO$_{5}$, and to previously published diffraction data for
Er$_{2}$BaNiO$_{5}$. From this analysis we extract the bare staggered
magnetization curve for  Haldane spin chains in these compounds.
\end{abstract}
\pacs{75.25.+z,75.10.-b,75.10.Jm,75.50.Ee}

\narrowtext
%\twocolumn
%====================================================================================
The quantum-disordered ground state and the famous Haldane energy gap in
the magnetic excitation spectrum\cite{Haldane83} have kept
one-dimensional (1-D) integer-spin Heisenberg antiferromagnets (HAF) at
the center of attention of condensed matter physicists for  the last 15
years. Among the more recent developments are studies of  such systems
in external uniform magnetic
fields\cite{Kakurai91,Kobayashi94,Regnault97}. It was found that in
sufficiently strong fields one of the three Haldane-gap modes undergoes
a complete softening at some particular wave vector. The result is a
transition to a new phase with long-range antiferromagnetic correlations
(see for example Refs.~\onlinecite{Golinelli,Sorensen93}). The effect of
a {\it staggered} field ${\bf H}_{\pi}$, to which a Haldane chain is
most susceptible,  is expected to be no less dramatic. Unfortunately,
this problem has been given much less attention in literature, simply
because such conditions are almost impossible to realize in an
experiment. The only chance of producing a magnetic field modulated on
the {\it microscopic} scale is to make use of some periodic modulation
that is intrinsic to the system under investigation. This can be, for
example, a structural modulation, as is the case in  NENP, one of the
best-known Haldane-gap compounds (Refs.~\onlinecite{Regnault94,Ma92} and
references therein). The $S=1$ chains in this material consist of
alternating crystallographically non-equivalent Ni$^{2+}$ ions with
slightly different gyromagnetic ratios. A weak effective staggered field
can thus be induced in NENP by applying a {\it uniform} external
field\cite{Mitra94}. Unfortunately, the effect of the staggered
component is obscured by the response of the system to the strong
uniform field itself\cite{Mitra94,Sakai94}.

A more direct approach is to use an intrinsic {\it magnetic} modulation
in a material that, in addition to integer-spin Heisenberg chains, has
other magnetic ions. Should the latter become ordered magnetically with
an appropriate propagation vector, they will project an effective
staggered exchange field on the Haldane spin chains. The magnitude of
the staggered field can be varied in an experiment indirectly, by
changing the temperature, and thus the magnitude of the magnetic order
parameter. By measuring the induced moment on the Haldane chains one can
hope to directly measure the {\it staggered magnetization curve for a
Haldane spin system}. This function, that  we shall denote as ${\cal
M}(H_{\pi})$, is one of the principal characteristics of a
quantum-disordered integer-spin HAF, but to date has not been determined
experimentally. In the present paper we shall demonstrate how ${\cal
M}(H_{\pi})$ can be determined by properly analyzing the results of
simple neutron diffraction experiments.

It appears that the best experimental realizations of the mechanism
described in the previous paragraph are to be found among linear-chain
nickelates with the general formula $R_{2}$BaNiO$_{5}$ ($R$=magnetic
rare earth). The $S=1$ Ni$^{2+}$ ions in these compounds are arranged in
distinct chains that run along the $a$ axis of the orthorhombic crystal
structure. In-chain interactions between the Ni spins are
antiferromagnetic, with the exchange constant $J\approx 300$~K, and
interchain coupling is negligible. The magnetic $R^{3+}$ ions are
positioned in between the Ni chains, and are expected to be weakly
coupled to the Ni$^{2+}$ moments (exchange  parameters of the order of
tens of Kelvin) and even weaker between themselves (coupling constants
of less than 1 K)\cite{Garcia95}. All $R_{2}$BaNiO$_{5}$ species order
antiferromagnetically with N\'{e}el temperatures ranging from 16 to
80~K\cite{Garcia95,Garcia97}. In the ordered phase a non-zero ordered
moment is found on both $R^{3+}$ and Ni$^{2+}$ sites. The magnetic order
parameter of Ni$^{2+}$ never achieves complete saturation, and at
$T\rightarrow 0$ is substantially smaller than the expected value of
$2\mu_{B}$ per ion, a hint at that quantum spin fluctuations in the
Ni-chains survive well below the N\'{e}el temperature $T_{N}$.  Haldane
gap excitations propagating on the $S=1$ Ni-chains have been observed
both above and below the N\'{e}el temperature in all materials studied
so far, including Pr$_{2}$BaNiO$_{5}$\cite{Zheludev96PBANO},
Nd$_{2}$BaNiO$_{5}$\cite{Zheludev96NBANO}, and
(Nd$_{x}$Y$_{1-x}$)$_{2}$BaNiO$_{5}$\cite{Yokoo97NDY}. These spin
excitations have a purely 1-D dispersion and dynamic structure factor,
and are practically indistinguishable from Haldane modes in
Y$_{2}$BaNiO$_{5}$, an extensively studied reference  Haldane gap
system, where no long-range magnetic ordering occurs even at low
temperatuures\cite{Darriet93,DiTusa94,Sakaguchi96,Xu96}. The basis of
our current understanding of $R_{2}$BaNiO$_{5}$ compounds is that the
Haldane singlet ground state of individual $S=1$ Ni-chains is preserved
in the magnetically ordered phase\cite{Zheludev96PBANO,MZ}.  The
non-zero ordered moment on the Ni$^{2+}$ sites is viewed as a result of
polarization of the  quantum-disordered  $S=1$ chains by an effective
staggered exchange field,  generated by the ordered $R$-sublattice.

Based on this physical picture, we shall construct a simple mean-field
(MF) model  of magnetic ordering in $R_{2}$BaNiO$_{5}$ materials. In the
conventional MF  approach  all magnetic ions  are treated as {\it
isolated} moments that become polarized by the effective exchange field.
In the $R_{2}$BaNiO$_{5}$'s however, in-chain Ni-Ni exchange  coupling
is dominant over all other magnetic interactions. It is therefore more
appropriate to treat only the  rare earths ions as individual
moments\cite{Garcia95}. In the spirit of what is said above, the bare
magnetization curve for the Ni-sublattice should be taken in the form of
the staggered magnetization function ${\cal M}$ for an isolated $S=1$
quantum spin chain. Ni-Ni interactions are {\it not} to be included
explicitly into the MF theory, since they are already built into the
function ${\cal M}$. The MF equations for the Ni-subsystem can thus be
written in the following form:
\begin{eqnarray}
 M^{({\rm Ni})}=gS\mu_{B} {\cal M}(H^{{\rm (Ni)}}),\label{e1}\\
 H^{{\rm (Ni)}}=2 \alpha_{R} M^{(R)}\label{e2}.
 \end{eqnarray}
Here $M^{({\rm Ni})}$ and $M^{(R)}$ are the sublattice magnetizations
(per ion) for the Ni and $R$ subsystems, respectively, $H^{{\rm (Ni)}}$
is the effective staggered field that acts on the Ni-chains, $g=2$ is
the gyromagnetic ratio for Ni$^{2+}$, $S=1$ is the spin of Ni$^{2+}$,
and $\alpha_{R}$ is the effective MF coupling constant. To complete the
model, we have to write down the MF equations for the $R$-subsystem as
well. Unfortunately, due to the low site-symmetry for $R^{3+}$ ions in
$R_{2}$BaNiO$_{5}$ materials\cite{Zheludev96PBANO}, modeling the exact
bare magnetization curve of the rare earths is impossible without
knowing the details of their electronic structure. Nevertheless, for
those compounds in which $R^{3+}$ is a Kramers ion, we can hope to get a
reasonable approximation using the Brillouin function:
\begin{eqnarray}
 M^{(R)}=M_{0}^{(R)}\tanh(H^{(R)} M_{0}^{(R)}/\kappa T),\label{e3}\\
 H^{(R)}=\alpha_{R} M^{({\rm Ni})}.\label{e4}
 \end{eqnarray}
Here $M_{0}^{(R)}$ is the effective moment of the rare earth ion and
$H^{(R)}$ is the mean field acting on the rare earths. Direct $R$-$R$
interactions are expected to be weak, and $T_{N}$  is predominantly
defined by the magnitude of $R$-Ni exchange. This fact enables us to
simplify the equations  by omitting $R$-$R$  coupling in Eq.~\ref{e4}.

Our ``semi-quantum''  model differs significantly from the usual MF
model for classical magnets. In the latter, at $T\rightarrow 0$ all
sublattices become fully saturated as the bare single-ion susceptibility
diverges as $1/T$. In contrast, in our model the bare staggered
susceptibility of the Haldane chains remains {\it finite} at $T=0$.
Moreover, at $T\lesssim \Delta$ ($\Delta \approx 9$~meV, or $\approx
105$~K $R_{2}$BaNiO$_{5}$ compounds), the function ${\cal M}(H_{\pi})$
is expected to be almost $T$-independent. Our model will therefore
produce qualitatively different results for the $T$-dependencies and
saturation values of the magnetic order parameter for Ni.

It has to be noted that the idea of treating interchain coupling at the
MF or RPA (Random Phase Approximation) level is, in itself, not a new
idea, but a well-established technique\cite{Scalapino75}. In their
pioneering work on CsNiCl$_{3}$, the first Haldane-gap material studied
experimentally, Buyers et al. \cite{BuyersCsNiCl3}, and Affleck
\cite{Affleck89} implemented this approach to explain magnetic ordering
and calculate the spin wave dispersion relations. The main difference
between CsNiCl$_{3}$ and $R_{2}$BaNiO$_{5}$ is that exchange coupling
between individual Haldane chains is {\it direct} in the former system,
and {\it mediated by the rare earth ions} in the latter. For directly
coupled Haldane spin chains, the magnitude of interchain interactions
must exceed some critical value in order for the system to order
magnetically \cite{Affleck89}. In a MF treatment, the temperature
dependence of the ordered moment is defined by the {\it intrinsic}
temperature dependence of the susceptibility of individual chains. In
contrast, in our case of $R_{2}$BaNiO$_{5}$ compounds, magnetic ordering
is driven by the $1/T$-divergent susceptibility of the rare earth
subsystem. At sufficiently low temperature long-range order will
therefore occur for arbitrary small $R$-Ni interactions, and  in the
case of $T_{N}\lesssim \Delta$, we can use the approximation where the
bare magnetization curve of isolated chains is $T$-independent.
$R_{2}$BaNiO$_{5}$ materials are  thus more convenient as model systems
for studying the effect of staggered field on Haldane spin chains.

To test the validity of our model we shall compare its predictions for
the temperature dependence of sublattice magnetizations with
experimental results obtained using neutron diffraction. Since one of
our primary goals  is to verify that the function ${\cal M}$ is indeed
$T$-independent at $T\lesssim
\Delta$, we need to compare experimental $M(T)$ curves measured in material with
substantially different ordering temperatures. Moreover, as the constant
$\alpha_{R}$ is clearly dependent of $R$, the only way to exclude $T$ as
an implicit variable is to study several dilute systems with the formula
($R_{x}$Y$_{1-x}$)$_{2}$BaNiO$_{5}$. In  such compounds part of the
magnetic $R^{3+}$'s are randomly replaced by non-magnetic Y$^{3+}$ ions.
$T_{N}$ is strongly dependent on the rare earth concentration
$x$\cite{Yokoo97NDY}. At the same time, we can expect the coupling
constant $\alpha_{R}$ to be the same in all
($R_{x}$Y$_{1-x}$)$_{2}$BaNiO$_{5}$ systems, provided in Eq.~\ref{e2}
$M^{(R)}$ is replaced by the average moment on the $R$ sites, i.e., by
$\tilde{M}^{(R)}=x M^{(R)}$. Indeed, in the crystal structure each Ni
site is coordinated to four $R$-sites, and any effect of disorder, for
not too small $R$-content, is reduced by averaging over nearest
neighbors, and the effective staggered field scales proportionately to
$x$.

To date, accurate $T$-dependent data exist only for Ho$_2$BaNiO$_{5}$
($T_{N}=53$~K)\cite{Garcia93} and Er$_2$BaNiO$_{5}$
($T_{N}=33$~K)\cite{Alonso90}.  Of these  materials only the Er
nickelate fits the  condition of having a Kramers rare earth ion,
essential for using Eq.~\ref{e3}. To have more compounds to work with,
and to apply our model to systems with substantially different
$R$-concentrations $x$, we have performed new powder neutron diffraction
studies of Nd$_{2}$BaNiO$_{5}$ ($T_{N}=48$~K, $x=1$) and NdYBaNiO$_{5}$
($T_{N}=29$~K, $x=0.5$). The experiments were done  on the position
sensitive detector diffractometer D1B at the Institut Laue-Langevein,
Grenoble, France. Details of the rather straightforward experiment and
data analysis will be reported elsewhere. At all temperatures the
obtained magnetic structures in both materials are defined by the
propagation vector $(1/2,0,1/2)$, and are similar to that in
Ho$_2$BaNiO$_{5}$\cite{Garcia93}. In NdYBaNiO$_{5}$ the Ni$^{2+}$ and
Nd$^{3+}$ moments lie in the $(a,c)$ crystallographic plane, and form
almost temperature-independent angles of $\phi^{({\rm Ni})}\approx
35^{\circ}$ and $\phi^{({\rm Nd})}\approx 0^{\circ} $ with the $c$-axis,
respectively. The same applies to Nd$_{2}$BaNiO$_{5}$  at $T\lesssim
40$~K. Between $40$ and $48$~K however, i. e., just below $T_{N}$, both
the Ni and Nd ordered moments in Nd$_{2}$BaNiO$_{5}$ undergo a
significant re-orientation, in agreement with Ref.\cite{Garcia97}. Our
principal experimental results, the temperature dependencies of the
ordered moments, are summarized in Fig.~\ref{mag}. In the same figure we
show the data for Er$_{2}$BaNiO$_{5}$, taken from Ref.~\cite{Alonso90}.
In Er$_{2}$BaNiO$_{5}$ the orientation of the ordered moments of
Ni$^{2+}$ and Er$^{3+}$ is $T$-independent, with $\phi^{({\rm
Ni})}=65^{\circ}$ and $\phi^{({\rm Er})}
=88^{\circ}$, respectively. In the Ho-system the orientation of ordered moments
change only slightly with decreasing temperature, and the angles are,
correspondingly, $\phi^{({\rm Ni})} \approx 25^{\circ}$ and $\phi^{({\rm
Ho})}\approx 0^{\circ}$ \cite{Garcia93}.

We are now in a position to plot $M^{({\rm Ni})}$ as a function of
$\tilde{M}^{(R)}$. For Nd$_{2}$BaNiO$_{5}$ and NdYBaNiO$_{5}$ ($x=1$ and
$x=0.5$, respectively) this curve is shown in Fig.~\ref{final} in solid
triangles and squares, correspondingly. For these two materials in which
the ordering temperatures differ by almost a factor of 2 we obtain an
excellent data collapse, which  confirms that the function ${\cal M}$ is
almost $T$-independent. To place the Er$_{2}$BaNiO$_{5}$ data on the
same plot we have to re-scale the absciss to compensate for the
difference between $\alpha_{{\rm Nd}}$ and $\alpha_{{\rm Er}}$. This is
done in such a way as to obtain the best overlap between the three
curves. The best data collapse is achieved by scaling down the Er moment
by a factor of 3.2(1) ( Fig.~\ref{final}, open circles). For
Ho$_{2}$BaNiO$_{5}$, a non-Kramers system, the approximation (\ref{e3})
is not appropriate, yet Eqs. (\ref{e1},\ref{e2}) are expected to remain
valid. We can therefore plot the existing data for Ho$_{2}$BaNiO$_{5}$
\cite{Garcia93} in the graph on Fig.~\ref{final}, applying the same
absciss-rescaling procedure as for Er$_{2}$BaNiO$_{5}$. The best data
collapse in this case is obtained by scaling down the Ho moment by a
factor 4.2(1) ( Fig.~\ref{final}, open diamonds). We see that, despite
the different $R$-substitute, the overall shape of the $M^{({\rm Ni})}$
vs. $\tilde{M}^{(R)}$ curve  in all four materials is very similar, in
support of our conjecture that ${\cal M}$ is an intrinsic property of
the Ni chains and is the same for all $R_{2}$BaNiO$_{5}$ systems. For
further use we shall fit the cumulative data  in Fig.~\ref{final} with a
purely empirical function:
\begin{equation}
gS\mu_{B}{\cal M}(\alpha_{R} \tilde{M}^{(R)})=A \arctan (B_{R}
\tilde{M}^{(R)}),\label{fun}
\end{equation}
A good fit (solid line in Fig.~\ref{final})  is obtained with
$A=1.17(4)~\mu_{B}$,  $B_{{\rm Nd}}=1.7(1)~\mu_{B}^{-1}$, $B_{{\rm
Er}}\equiv B_{{\rm Nd}}/3.2=0.43(3)~\mu_{B}^{-1}$, and $B_{{\rm
Ho}}\equiv B_{{\rm Nd}}/4.2=0.33(3)~\mu_{B}^{-1}$.

 In the
context of our model the curve in Fig.~\ref{final} is nothing else but
the staggered magnetization function for a Haldane spin chain that we
are trying to determine in this study. The only  thing we are missing at
this point is a proper scale (units of magnetic field) on the absciss,
which is currently labeled in units of magnetization. Conversion to
field units is done by the coupling constant $\alpha$, that   we shall
obtain by analyzing the temperature-dependent experimental data with our
mean-field equations.  The analysis  can be  performed for $R=$Nd/Y and
$R=$Er, but not for $R=$Ho, where, as discussed above, Eq.~\ref{e3} does
not apply. In place of $M_{0}^{(R)}$ in Eq.~\ref{e4} we shall use the
saturation moment for the rare earth ions in each system. That the
values of $M_{0}^{({\rm Nd})}$ are slightly different in the two
Nd-based compounds ($2.68\mu_{B}$ and $2.3\mu_{B}$,respectively) only
emphasizes the fact that Eq.~\ref{e4} is no more than an approximation
to the actual magnetization function for an isolated $R^{3+}$ ion.
Combining our MF equations with Eq.~\ref{fun} we use least-squares
refinement to fit the experimental data in Fig.~\ref{mag}, having
$\alpha_{R}$ as the only adjustable parameter for each compound. Very
good fits  are obtained for all systems (solid lines in Fig.~\ref{mag}).
The only discrepancy is  above 40~K for Nd$_{2}$BaNiO$_5$, where the
magnetic structure changes dramatically and our model is not applicable
anyway. The refined values for the coupling constants are: $\alpha_{{\rm
Nd}}^{(1)}=0.82(2)\cdot 10^{5}$~Oe/$\mu_{B}$ in Nd$_{2}$BaNiO$_5$,
$\alpha_{{\rm Nd}}^{(2)}=0.99 (2)\cdot 10^{5}$~Oe/$\mu_{B}$ in
NdYBaNiO$_5$, and $\alpha_{{\rm Er}}=0.183(3)
\cdot 10^{5}$~Oe/$\mu_{B}$ in Er$_{2}$BaNiO$_5$. Obtaining similar
values of $\alpha_{{\rm Nd}}$ for the two Nd-based  nickelates is
another self-consistency check for our approach. Equation \ref{e2} can
now be used to replot the induced moment on the Ni-chains as a function
of effective exchange field.  As the data collapse in Fig.~\ref{final}
is rather good,  for this purpose we can use the effective average value
$\overline{\alpha}_{{\rm Nd}}\equiv(\alpha_{{\rm Nd}}^{(1)}+\alpha_{{\rm
Nd}}^{(2)}+3.2
\alpha_{{\rm Er}})/3= 0.8(1)\cdot 10^{5}$~Oe/$\mu_{B}$. The resulting staggered field scale is
shown in the top axis in Fig.~\ref{final}.

Our model, simple-minded as it is, appears to adequately describe the
magnetic ordering in all $R_{2}$BaNiO$_5$ systems studied so far. Its
application provides the first measurement of the staggered
magnetization curve for a Haldane spin chain. We have recently succeeded
in calculating this function theoretically, extending the approach laid
out in Ref. \cite{MZ}. This calculation is in good agreement with our
experimental data, but the discussion is beyond the scope of this paper
and will be addressed in a forthcoming publication.

This study was in part supported by the U.S.-Japan Cooperative Program
on Neutron Scattering, a Grant-in-Aid for Scientific Research from the
Ministry of Education, Science and Culture Japan and The Science
Research Fund of Japan Private School Promotion Foundation. Work at
Brookhaven National Laboratory was carried out under Contract No.
DE-AC02-76CH00016, Division of Material Science, U.S.\ Department of
Energy.

%\bibliographystyle{prsty}
%\bibliography{nhaldane}

\begin{thebibliography}{10}

\bibitem{Haldane83}
F.~D.~M. Haldane, Phys. Rev. Lett. {\bf 50},  1153  (1983).

\bibitem{Kakurai91}
K. Kakurai, M. Steiner, R. Pynn, and J.~K. Kjems, J. Phys.: Condens. Matter
  {\bf 3},  715  (1991).

\bibitem{Kobayashi94}
T. Kobayashi {\it et~al.}, J. Phys. Soc. Jpn. {\bf 63},  1961  (1994).

\bibitem{Regnault97}
L.~P. Regnault and J.~P. Renard, Physica B {\bf 243-236},  541  (1997).

\bibitem{Golinelli}
O.~Golinelli, Th.~Jolicoeur and R.~Lacase, Phys. Rev. B {\bf 45}, 9798 (1992);
  J.~Phys.: Condens. Matter {\bf 5}, 7847 (1993).

\bibitem{Sorensen93}
E.~S. Sorensen and I. Affleck, Phys. Rev. Lett. {\bf 71},  1633  (1993).

\bibitem{Regnault94}
L.~P. Regnault, I. Zaliznyak, J.~P. Renard, and C. Vettier, Phys. Rev. B {\bf
  50},  9174  (1994).

\bibitem{Ma92}
S. Ma {\it et~al.}, Phys. Rev. Lett. {\bf 69},  3571  (1992).

\bibitem{Mitra94}
P.~P. Mitra and B.~I. Halperin, Phys. Rev. Lett. {\bf 72},  912  (1994).

\bibitem{Sakai94}
T. Sakai and H. Shiba, J. Phys. Soc. Jpn. {\bf 63},  867  (1994).

\bibitem{Garcia95}
E. Garc\'{i}a-Matres, J.~L. Garc\'{i}a-Munoz, J.~L. Mart\'{i}nez, and J.
  Rodriguez-Carvajal, J. Mag. Magn. Mat. {\bf 149},  363  (1995).

\bibitem{Garcia97}
E. Garc\'{i}a-Matres, J.~L. Mart\'{i}nez, and J. Rodriguez-Carvajal, Physica B
  {\bf 234--236},  567  (1997).

\bibitem{Zheludev96PBANO}
A. Zheludev, J. M. Tranquada, T. Vogt, D. J. Buttrey, Europhys. Lett. {\bf 35},
  385 (1996); Phys. Rev. B {\bf 54}, 6437 (1996).

\bibitem{Zheludev96NBANO}
A. Zheludev, J.~M. Tranquada, T. Vogt, and D.~J. Buttrey, Phys. Rev. B {\bf
  54},  7216  (1996).

\bibitem{Yokoo97NDY}
T. Yokoo, A. Zheludev, M. Nakamura, and J. Akimitsu, Phys. Rev. B {\bf 55},
  11516  (1997).

\bibitem{Darriet93}
J. Darriet and L.~P. Regnault, Solid State Commun. {\bf 86},  409  (1993).

\bibitem{DiTusa94}
J.~F. DiTusa {\it et~al.}, Physica B {\bf 194-196},  181  (1994).

\bibitem{Sakaguchi96}
T. Sakaguchi, K. Kakurai, T. Yokoo, and J. Akimitsu, J. Phys. Soc. Jnp
{\bf 65}, 3025 (1996).


\bibitem{Xu96}
G. Xu {\it et~al.}, Phys. Rev. B {\bf 54},    (1996).

\bibitem{MZ}
S. Maslov and A. Zheludev, Phys. Rev. B {\bf 57},  68  (1998).


\bibitem{Scalapino75}
D. J. Scalapino, Y. Imry and P. Pincus, Phys. Rev. B {\bf 11}, 2042
(1975).

\bibitem{BuyersCsNiCl3}
W. J. L. Buyers {\it et~al.}, Phys. Rev. Lett. {\bf 56} 371 (1986);
Morra {\it et~al.}, Phys. Rev. B {\bf 38}, 543 (1988) and references
therein.

\bibitem{Affleck89}
I. Affleck, Phys. Rev. Lett. {\bf 62}, 474 (1989).



\bibitem{Garcia93}
E. Garc\'{i}a-Matres, J. Rodr\'{i}guez-Carjaval, J.~L. Mart\'{i}nez, and A.
  Salinas-S\'{a}nchez, Solid State Comm. {\bf 85},  553  (1993).

\bibitem{Alonso90}
J.~A. Alonso {\it et~al.}, Solid State Comm. {\bf 76},  467  (1990).



\end{thebibliography}

\begin{figure}
\caption{Measured temperature dependencies of the Ni- (open symbols) and $R$-
(solid symbols) ordered moments in Nd$_{2}$BaNiO$_5$ and NdYBaNiO$_5$
(this work), and Er$_{2}$BaNiO$_5$ (digitized from
Ref.~\protect\onlinecite{Alonso90}). The solid lines are
single-parameter fits using the mean field model described in the text.}
\label{mag}
\end{figure}

\begin{figure}
\caption{Ordered  staggered moment on the Ni-chains plotted against
the ordered moment of the $R$-sublattice (bottom axes). The  solid line
is an empirical fit, as described in the text. The data for
Ho$_{2}$BaNiO$_5$ and Er$_{2}$BaNiO$_5$ are taken from Refs.\
\protect\cite{Garcia93,Alonso90}. The top axis shows the corresponding staggered exchange field
acting on the Ni-chains.}
\label{final}
\end{figure}

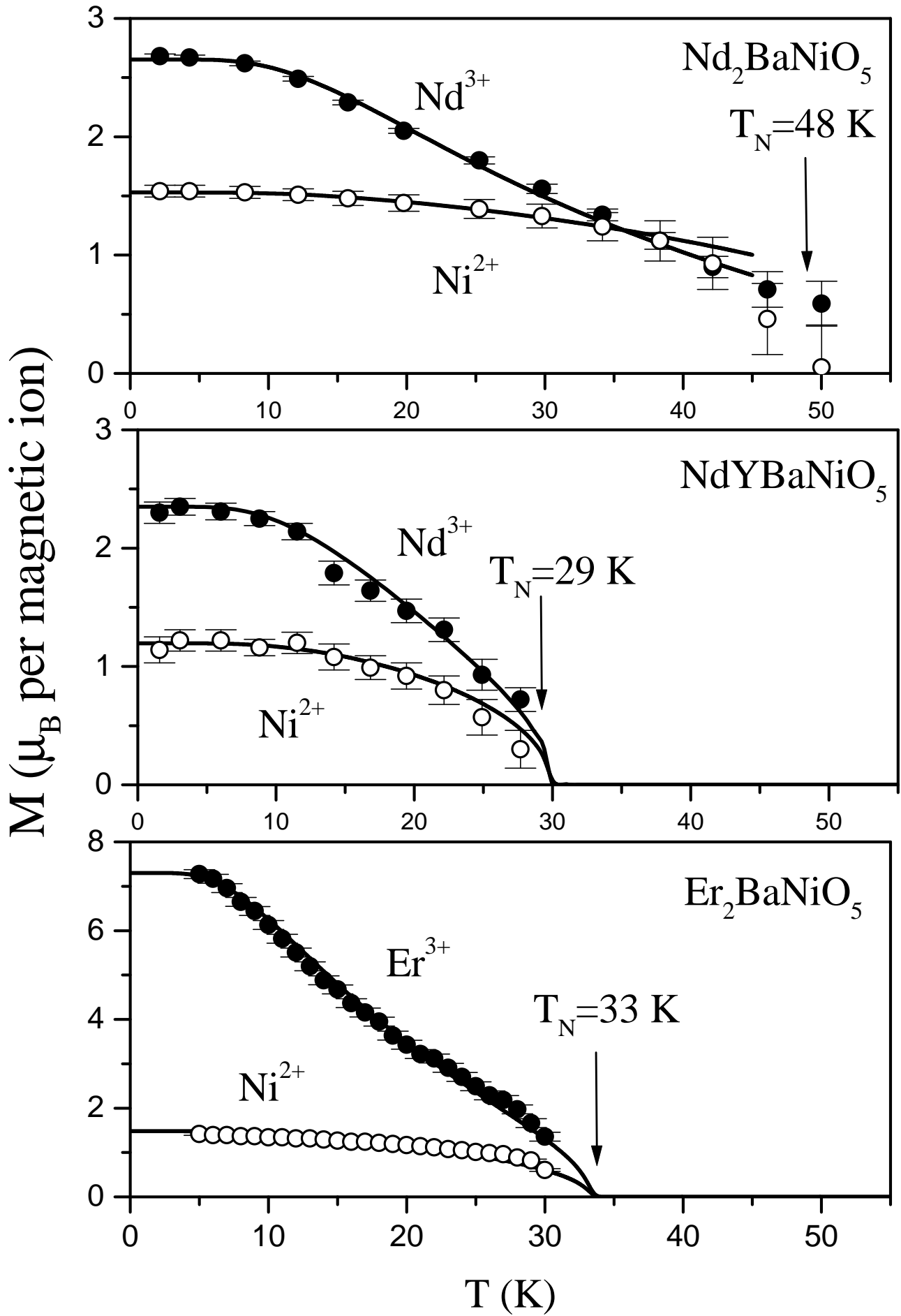
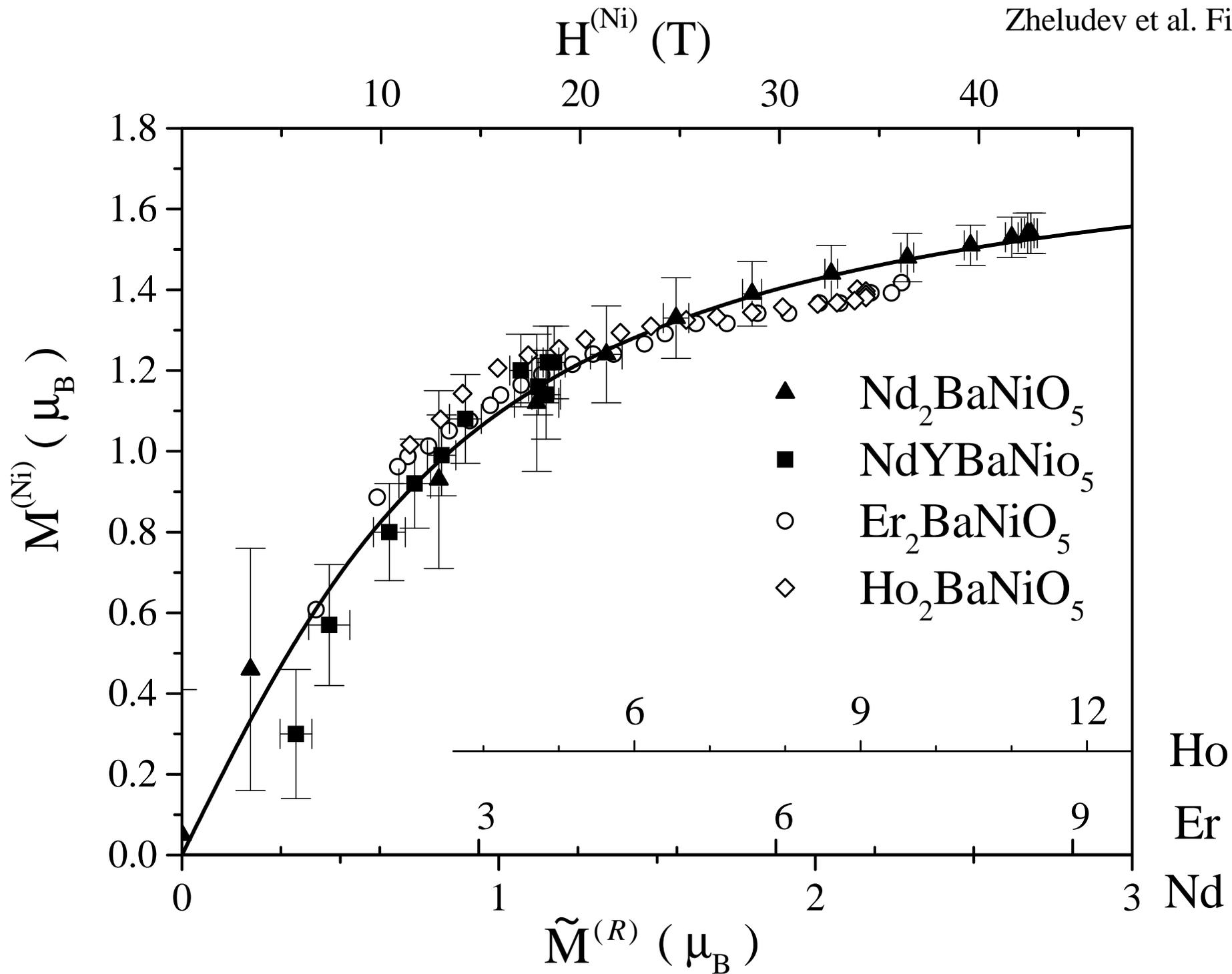
\end{document}